\documentclass[]{elsart}

\usepackage[]{graphicx}
\usepackage{subfigure}
\usepackage[latin1]{inputenc}
\usepackage{amsmath}
\usepackage[sort]{cite}
\usepackage{xspace}
\usepackage{amssymb}
\usepackage{hyperref}
\usepackage{multicol}
\newcommand{\ie}{{i.e.}\@\xspace}

\begin{document}
\begin{frontmatter}
	\title{The vacancy - edge dislocation interaction in fcc metals:
		a comparison between atomic simulations and elasticity theory
		}
	\author{Emmanuel Clouet},
	\ead{emmanuel.clouet@cea.fr}
	\address{Service de Recherches de Métallurgie Physique, CEA/Saclay, \\
	91191 Gif-sur-Yvette, France}
	\begin{abstract}
		The interaction between vacancies and edge dislocations 
		in face centered cubic metals (Al, Au, Cu, Ni) is studied 
		at different length scales.
		Using empirical potentials and static relaxation, atomic 
		simulations give us a precise description of this interaction,
		mostly in the case when the separation distance between
		both defects is small. 
		At larger distances, elasticity theory can be used to predict this 
		interaction. From the comparison between both approaches 
		we obtain the minimal separation distance where elasticity 
		applies and we estimate the degree of refinement required in the calculation. 
		In this purpose, isotropic and anisotropic elasticity is used
		assuming a perfect or a dissociated edge dislocation
		and considering the size effect as well as the inhomogeneity interaction.
	\end{abstract}
\end{frontmatter}

\section{Introduction}
Understanding the interaction between dislocations and vacancies is a key point 
in material science. In metal plasticity, this interaction controls
dislocation climb and thus the plastic strain of a metal at high temperature 
when the glide of dislocations is blocked by obstacles like precipitates or the dislocation forest.
This interaction also plays an important role in phase transformations.
Indeed, as dislocations act as sinks and sources for vacancies,
they control the way vacancies adjust their concentration in non-isothermal kinetics
and thus the diffusion. Moreover, the vacancy super-saturation along dislocations
leads to an enhancement of diffusion in their vicinity: dislocations behave thus as 
short-circuits, a phenomenon known as pipe diffusion.

The interaction between vacancies and dislocations can be modeled using elasticity theory
(for a review, see Ref.~\citen{BUL70}). In 1949, Cottrell and Bilby calculated the so-called
size contribution, showing that the interaction energy is equal to the pressure
created by the dislocation times the relaxation volume due to the vacancy \cite{COT49}.
This energy is thus decreasing with the inverse of the distance between both defects.
In 1961, Bullough and Newman applied results of Eshelby theory \cite{ESH57,ESH61}
and brought out another contribution, the inhomogeneity interaction \cite{BUL63}.
This contribution is proportional to the square of the pressure and of the shear stress
created by the dislocation and is always attractive. It is much shorter range 
than the size contribution as it is decreasing with the square of the inverse of the distance.
Such results are based on a continuum description of matter and assumes
that a vacancy can be modeled by an inclusion having different elastic constants
from the matrix.

First atomic calculations of the interaction between vacancies and dislocations
\cite{BUL68,ING70,PER72,MIL81,FAS83,YAK85} were used to check the validity of the elasticity theory
and to obtain results in the vicinity of the dislocation core where a continuum description 
breaks down.
For face centered cubic (fcc) metals\cite{PER72,YAK85}, 
the vacancy binding energy was found to reach a maximum 
in the region of the stacking fault rather than in the immediate vicinity of the partial dislocations. 
More surprisingly, the vacancy was found to be attracted to a dissociated
edge dislocation both above and below its glide plane.
Some authors interpreted that as a manifestation of the inhomogeneity interaction \cite{PER72}.
In the same way, for body centered cubic metals, 
an anomalous positive binding energy was found in some cases
on the tensile side of the dislocation elastic field \cite{BUL68,ING70} 
and the comparison with elastic
models show discrepancies even for large separation distances \cite{BUL68,FAS83}.
Most of these simulations have been performed before the development in the mid-eighties of reliable 
empirical potentials for metals, like the embedded atom method (EAM) \cite{DAW84}
or the tight-binding second-moment approximation (TB-SMA) \cite{FIN84,ROS89}. 
They were based on pair potentials which fail to correctly predict
elastic behavior. 
In particular, all cubic crystals modeled with pair potentials have to obey the Cauchy relation,
which restricts the elastic constants $C_{12}$ and $C_{44}$ to be equal. 
Therefore, one can legitimately wonder whether results obtained with these potentials 
are not biased.
On the other hand, EAM and TB-SMA potentials do not suffer such restrictions 
and have been shown to be well suited to study dislocation behavior.
Some authors used these potentials to study the behavior of vacancies 
in the vicinity of dislocations, in particular to simulate pipe-diffusion in fcc metals
\cite{HUA89,HUA91,HAK90,BOE96,HOA98,FAN00}.
These studies showed that the most stable position of the vacancy lies at the edge
of the supplementary half plane corresponding to a partial dislocation 
and that diffusion is faster along dislocations than in the bulk.
Nevertheless, no careful comparison between atomic simulations and elasticity theory 
was performed.
Such a comparison is useful to determine the range of validity of a continuum representation 
and the degree of refinement needed by elasticity.
Indeed, not all materials properties can be modeled at the atomic scale and 
a description of the behavior of the material intrinsic defects at a larger scale 
is still needed. It is particularly true for plasticity where properties 
are linked to the collective behavior of dislocations which has to be modeled
at a larger scale than the atomic one.
It is thus interesting to see what conclusions can be drawn at the atomic scale
with empirical potentials and how these results compare with predictions of the elasticity theory.

In the present work, we study the interaction between vacancies and dislocations
in fcc metals at the atomic scale using EAM potentials 
and at a mesoscopic scale with the elasticity theory. We limit the discussion to edge dislocations
as creep is controlled by their climb and thus by their interaction with vacancies
when an external stress is applied.
Several metals having the fcc structure are studied so as to bring out general results 
for fcc metals and identify a possible influence 
of the material properties, like the stacking fault energy or the elastic behavior.
The results presented in this article are directly obtained from atomic simulations
and then compared with predictions of the elasticity theory using different levels 
of approximation.

\section{Atomic simulations}
\subsection{Interatomic potentials}

\begin{table}[!bp]
	\centering
	\caption{Properties of the different EAM potentials:
	lattice parameters $a$,
	vacancy energy of formation $E^{\mathrm{f}}_{\mathrm{V}}$,
	vacancy relaxation volume $\delta\Omega_{\mathrm{V}}$,
	stacking fault energy $\gamma$, 
	elastic constants $C_{11}$, $C_{12}$, and $C_{44}$,
	elastic anisotropy factor $A$,
	isotropic shear modulus $\mu$ 
	and Poisson coefficient $\nu$
	given by Voigt average (Eq.~\ref{eq:elastic_Voigt}),
	and dissociation distance as given by isotropic ($d^{\mathrm{iso}}$) 
	and anisotropic ($d^{\mathrm{ani}}$) elastic theories
	and atomic simulations ($d^{\mathrm{ato}}$).}
	\label{tab:para}
	\begin{tabular}{lrrrrr}
		\hline
		Potential		& Al \cite{ERC94}	& Au \cite{MED01}	& Cu \cite{MIS01}	& Ni \cite{ANG95,BAS97} \\
		\hline
		$a$ (\AA)		&	4.032		& 4.079			&	3.615		&	3.52		\\
		$E^{\mathrm{f}}_{\mathrm{V}}$ (eV) & 0.693	& 0.873			&	1.272		&	1.592		\\
		$\delta\Omega_{\mathrm{V}}$ (\AA$^3$) & -6.6	& -10.0			&	-3.5		&	-1.9		\\
		$\gamma$ (mJ.m$^{-2}$)	&	104		& 30.7			&	44		&	89		\\
		$C_{11}$ (GPa)		&	118		& 185.8			&	170		&	246		\\
		$C_{12}$ (GPa)		&	62.3		& 157.1			&	123		&	147		\\
		$C_{44}$ (GPa)		&	36.7		& 38.9			&	76.2		&	125		\\
		$A=2C_{44}/(C_{11}-C_{12})$ &	1.32		& 2.71			&	3.2		&	2.5		\\
		$\mu$ (GPa)		&	33.2		& 29.1			&	55.2		&	94.7		\\
		$\nu$			&	0.320		& 0.418			&	0.324		&	0.277		\\
		$d^{\mathrm{iso}}$ (\AA) &	11.7		& 43.4			&	37.0		&	27.5		\\
		$d^{\mathrm{ani}}$ (\AA) &	11.7		& 41.7			&	35.0		&	26.7		\\
		$d^{\mathrm{ato}}$ (\AA) &	15		& 40			&	31		&	25		\\
		\hline
	\end{tabular}
\end{table}

The atomic simulations in this work are performed using empirical potentials 
of the EAM type.
Such potentials are widely used to study dislocation properties
as they present a good compromise between the size of the system 
which can be simulated and the realism of dislocation simulations.
So as to test the generality of the results, various metals
with the fcc structure are examined.
The corresponding potentials are 
the ones developed by Ercolessi and Adams for Al\cite{ERC94},
by Medlin \etal{ } for Au\cite{MED01},
by Mishin \etal{ }for Cu \cite{MIS01},
and by Angelo \etal{ }for Ni \cite{ANG95}.
Some potentials leading to a better description of these metals properties 
may exist in the literature, 
but the purpose of this study is mainly to draw the main features
of the vacancy-dislocation interaction which are common to fcc metals
and to test the validity of the elastic theory compared to atomic simulations.
In this purpose, the chosen EAM potentials appear to be well-suited
as they cover a wide range of properties, as summarized in Tab.~\ref{tab:para}. 
For instance, Au potential leads to a low stacking fault energy $\gamma$ 
and Al potential to a high one.
Concerning their elastic behavior, Al potential appears to be almost isotropic
($A\sim1$) whereas Cu potential is strongly anisotropic. 
Moreover, Al potential gives quite soft elastic constants whereas Ni potential gives hard ones.
At last, the vacancy energy of formation obtained in Al is low compared to the one in Ni.

\subsection{Simulation cell}

\begin{figure}[!bp]
	\begin{minipage}{0.47\textwidth}
		\subfigure[]{\includegraphics[width=0.99\linewidth]{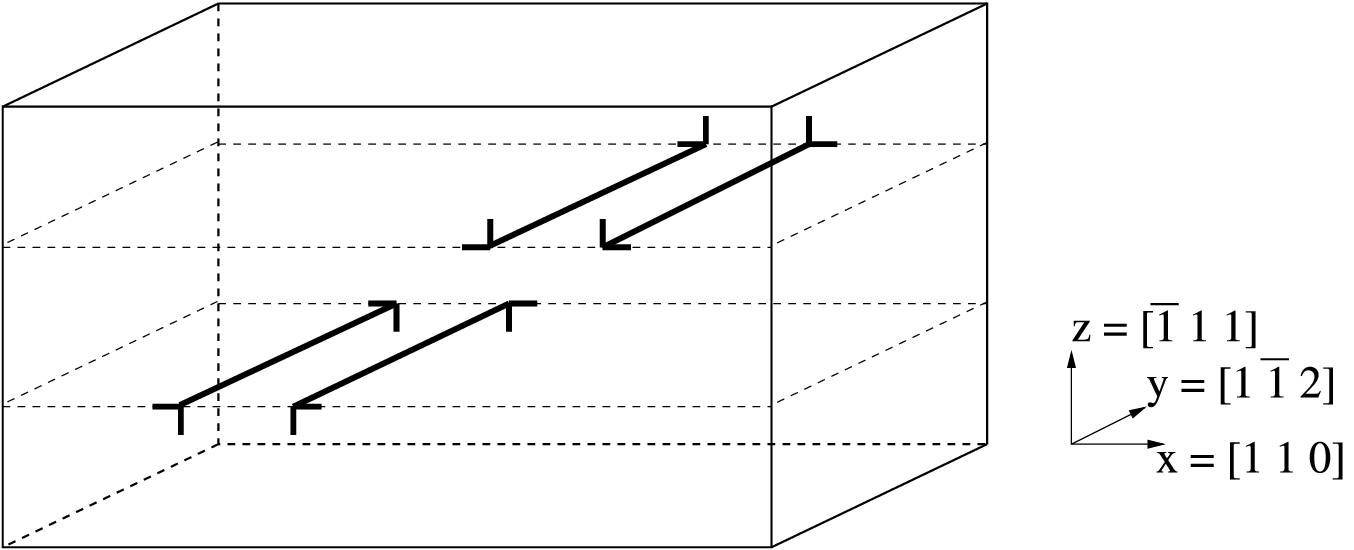}}
	\end{minipage}
	\hfill
	\begin{minipage}{0.47\textwidth}
		\subfigure[]{\includegraphics[angle=270,width=0.99\linewidth]{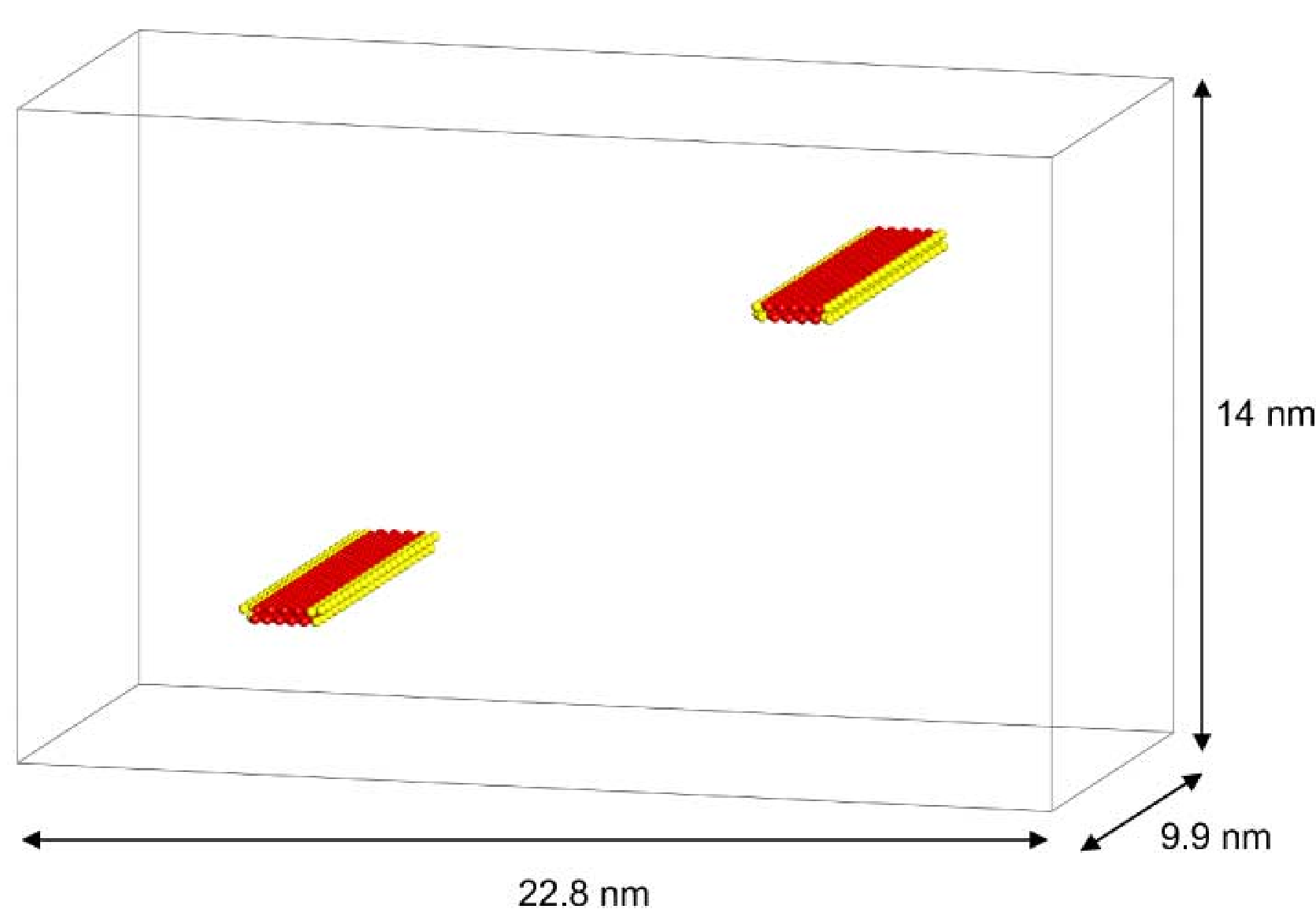}}
	\end{minipage}
	\caption{(a) Sketch of the simulation cell showing the two dissociated edge dislocations
	and their glide plane.
	(b) Aluminum simulation box containing two edge dislocations. For clarity, only atoms 
	belonging to the dislocations are shown: atoms in yellow (clear) are in the 
	core of the partial dislocations and atoms in red (dark) in the stacking fault.
	The cell contains respectively 160, 120 and 60 in $x$, $y$ and $z$ directions, 
	corresponding to 190800 atoms.}
	\label{fig:box_sketch}
\end{figure}

The main characteristics of the unit cell used for atomic 
simulations are adapted from those used in \cite{ROD99,ROD00,ROD04B}
and are sketched in Fig.~\ref{fig:box_sketch}a.
The $x$, $y$, and $z$ directions are parallel, respectively, 
to the $[110]$, $[1\bar{1}2]$, and $[\bar{1}11]$ directions.
With such orientations, perfect edge dislocations can easily be
introduced with a glide plane lying between two $(\bar{1}11)$ planes,
a line direction along the $[1\bar{1}2]$ axis 
and a Burgers vector along the $[110]$ axis.
As periodic boundary conditions are used in all directions, 
two edge dislocations with opposite Burgers vectors have to be 
introduced in the simulation box. 
The distance between the two corresponding glide planes
is chosen as half the height of the simulation box. 
According to elasticity theory, these two interacting dislocations
form a stable dipole for a $\pi/4$ angle between
the vector linking the centers of the dislocation lines
and the glide planes \cite{FRI64}.
This result holds for an isolated dipole, but, 
due to periodic boundary conditions, the dipole in the simulation box
interacts with an infinity of dipoles. 
We thus choose to put a distance equal to the half of the unit cell dimension 
in the $[110]$ direction and we find this position to be stable.
One should notice that dislocations in the simulation box are
straight dislocations and do not have any kink nor jog.

The atomic positions and the shape of the unit cell are relaxed at 0~K
using a conjugate gradient algorithm. 
As expected, each edge dislocations dissociates into two partial dislocations
separated by a stacking fault \cite{FRI64,HIR82}.
Atoms belonging to the core of these partial dislocations do not 
show fcc nor hexagonal compact neighborhood. They can therefore be easily identified
in the simulation box (Fig.~\ref{fig:box_sketch}b), allowing
the ``measurement'' of the dislocation separation distance $d^{\mathrm{ato}}$.
In agreement with the variations of the stacking fault energies $\gamma$, 
this separation distance is small for Al, big for Au 
and in-between for Cu and Ni (Tab.~\ref{tab:para}).
The convergence of the dissociation distance $d^{\mathrm{ato}}$
with the dimensions of the unit cell has been checked and
is found to be good for stackings of 160 and 60 planes
respectively in $[110]$ and $[\bar{1}11]$ directions

\subsection{Vacancy-dislocation interaction}

\begin{figure}[!bp]
	\begin{center}
		\includegraphics[width=0.99\textwidth]{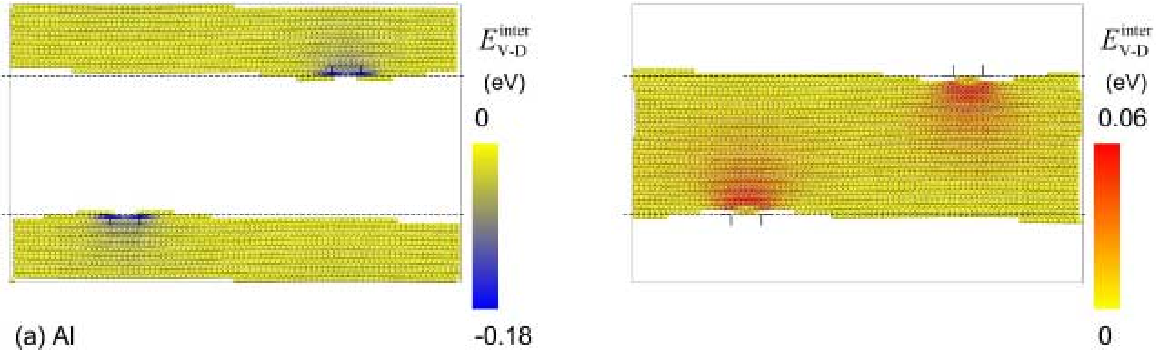}

		\vspace{1cm}
		
		\includegraphics[width=0.99\textwidth]{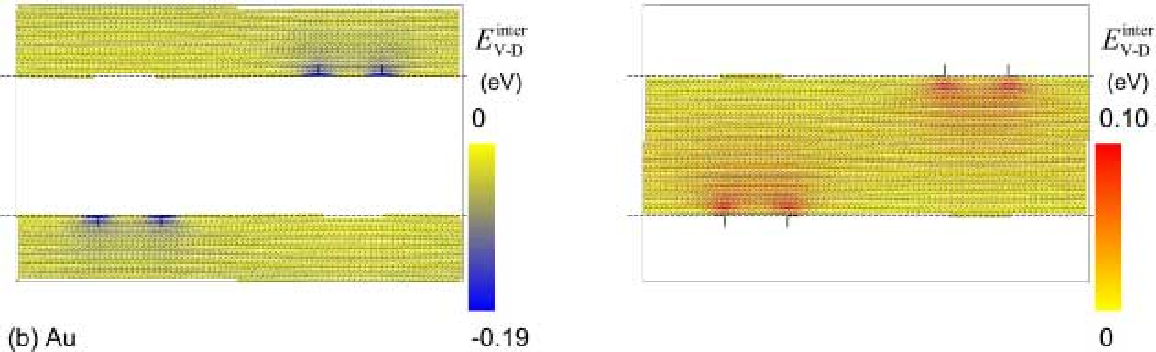}

		\vspace{1cm}

		\includegraphics[width=0.99\textwidth]{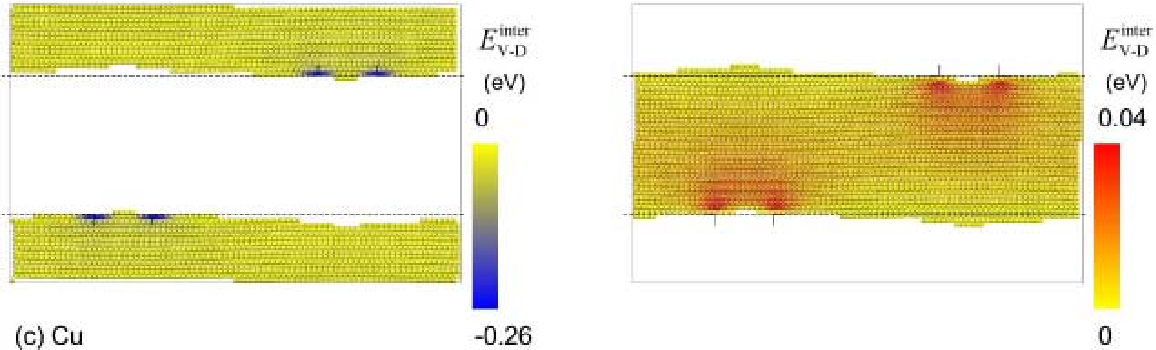}

		\vspace{1cm}

		\includegraphics[width=0.99\textwidth]{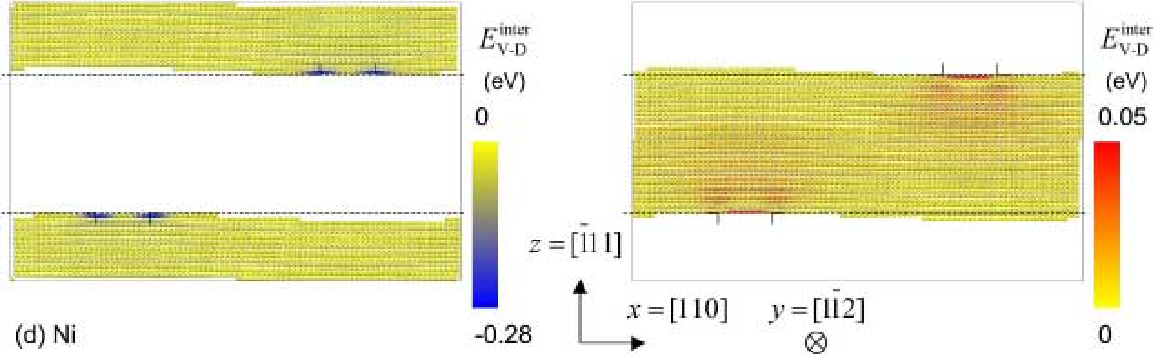}
	\end{center}
	\caption{Vacancy-dislocation interaction energy $E^{\mathrm{inter}}_{\mathrm{V-D}}$
		in (a) Al, (b) Au, (c) Cu, and (d) Ni as given by EAM potentials.
		On the left part of the figure, only attractive positions for the vacancy 
		are shown ($E^{\mathrm{inter}}_{\mathrm{V-D}}<0$), whereas on the right part
		only repulsive positions are shown ($E^{\mathrm{inter}}_{\mathrm{V-D}}>0$).
		The positions of the partial dislocations and of their glide planes are sketched by lines.}
	\label{fig:vac_dislo_inter_ato}
\end{figure}

\begin{figure}[!bp]
	\begin{center}
		\subfigure[Al]{\includegraphics[width=0.49\linewidth]{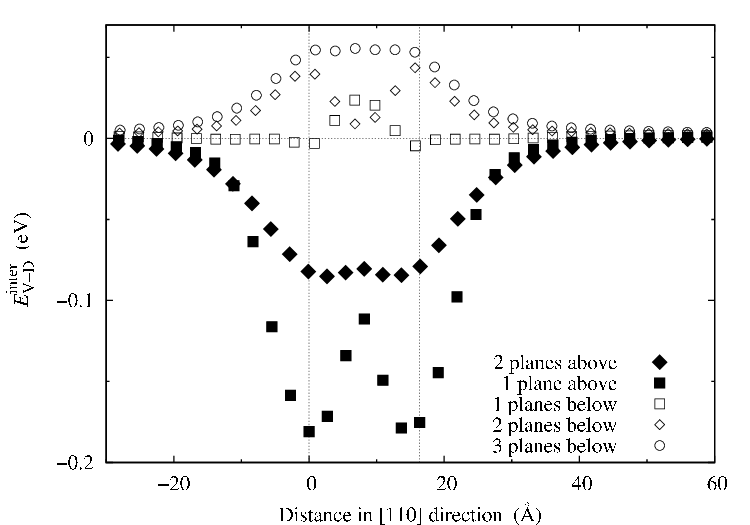}}
		\subfigure[Au]{\includegraphics[width=0.49\linewidth]{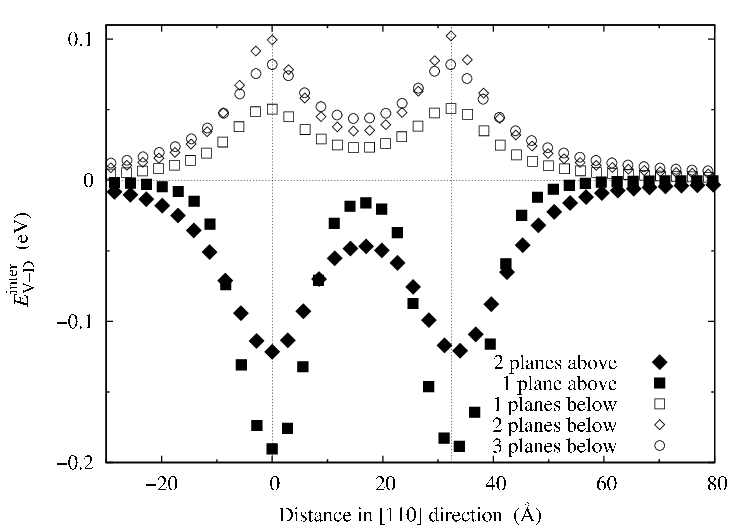}}
		\subfigure[Cu]{\includegraphics[width=0.49\linewidth]{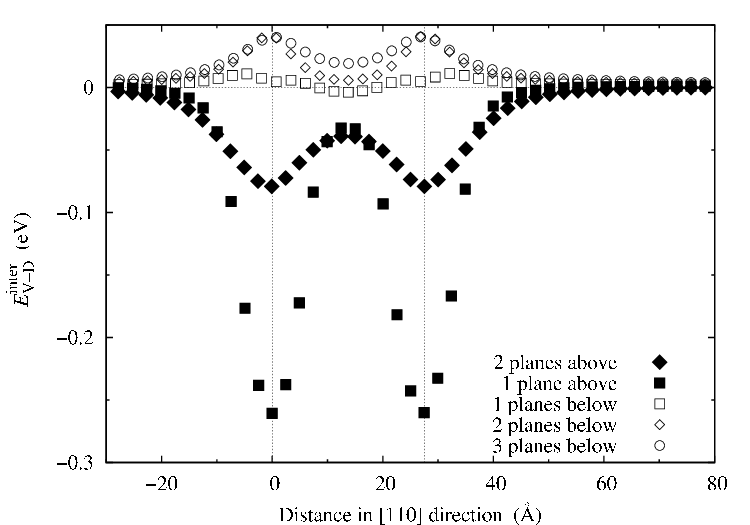}}
		\subfigure[Ni]{\includegraphics[width=0.49\linewidth]{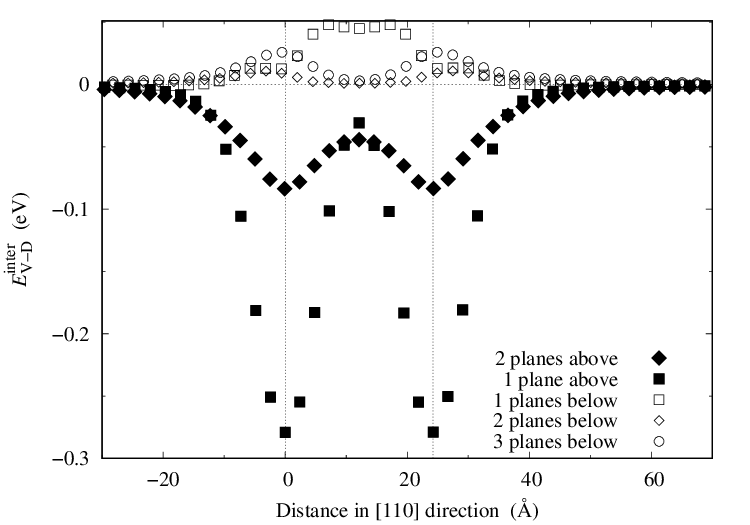}}
	\end{center}
	\caption{Vacancy-dislocation interaction energy $E^{\mathrm{inter}}_{\mathrm{V-D}}$
		in (a) Al, (b) Au, (c) Cu, and (d) Ni as a function of distance from the center 
		of one of the two partial dislocations. 
		Scans are performed in the [110] direction for different $(\bar{1}11)$ planes
		lying above (in the compressive zone) or below the glide plane (in the tensile zone).
		The origin is defined as the center of one of the two partial dislocations
		and axes are defined on Fig.~\ref{fig:box_sketch}a.}
	\label{fig:vac_dislo_inter_linescan}
\end{figure}

\begin{table}[!bp]
	\centering
	\caption{Lowest vacancy-dislocation interaction energy $E^{\mathrm{inter}}_{\mathrm{V-D}}$ (in eV)
		corresponding to a vacancy located on the edge of a partial dislocation.}
	\label{tab:EinterVD}
	\begin{tabular}{lcccc}
		\hline
		Authors					& Al	& Au	& Cu	& Ni	\\
		\hline
		Present work				& -0.181& -0.190& -0.261& -0.279\\
		Huang \etal \cite{HUA89}		& 	&	& -0.37	&	\\
		Huang \etal \cite{HUA91}		&	&	& -0.27	&	\\
		Häkkinen \etal \cite{HAK90}		& 	&	& -0.14	&	\\
		von Boehm \etal \cite{BOE96}		&	& -0.11	& -0.22	&	\\
		\hline
	\end{tabular}
\end{table}

Using the different EAM potentials, we calculate the vacancy-dislocation 
interaction energy for all possible positions of the vacancy in the simulation box
with respect to the dislocations.
This energy is defined as 
\begin{equation}
	\label{eq:Einter}
	E^{\mathrm{inter}}_{\mathrm{V-D}} = 
	E + E_{\mathrm{V-D}} - E_{\mathrm{D}} - E_{\mathrm{V}} ,
\end{equation}
where $E$, $E_{\mathrm{V-D}}$, $E_{\mathrm{D}}$, and $E_{\mathrm{V}}$
are the energy of the simulation box without any defect,
with both the vacancy and the dislocations, with the dislocations only 
and with the vacancy only, respectively.
All these energies are calculated after relaxing 
the atomic positions and the shape of the unit cell.

The calculations results are displayed both as functions of the vacancy 
positions (Fig.~\ref{fig:vac_dislo_inter_ato})
and as energy profiles along lines parallel to the glide plane 
(Fig.~\ref{fig:vac_dislo_inter_linescan}). 
As expected, compressive regions attract the vacancy 
whereas tensile zones are repulsive. 
For potentials presenting a low stacking fault energy, an attractive
and a repulsive basin can be associated with each one of the partial 
dislocations, as they are well dissociated. For Al it is not so clear 
since the separation distance between the two partial dislocations is smaller.

For all potentials, the most attractive position is located in the core of the partial dislocations,
on the edge of the supplementary half-plane above the glide plane in the compressive region. 
This is in disagreement with first results obtained with pair potentials \cite{PER72,YAK85}
but confirms most recent studies using most reliable empirical potentials 
like EAM \cite{HAK90}, TB-SMA \cite{BOE96}, or pseudopotentials \cite{HUA89,HUA91}.
The corresponding interaction energies are displayed in Tab.~\ref{tab:EinterVD} 
and compared with previous calculations, showing a dependency on
the potential used for a given metal.
It seems that the higher the vacancy formation energy, 
the stronger the binding to the dislocation.
Using the same potential for Al, Hoagland \etal \cite{HOA98}
calculated the interaction energy of a vacancy with a screw dislocation
($E^{inter}_{\mathrm{V-D}}=-0.188$~eV)
and Picu and Zang \cite{PIC04} with a 60° dislocation ($E^{inter}_{\mathrm{V-D}}=-0.163$~eV).
The values obtained in these works are close to the one we find for an edge dislocation
($E^{inter}_{\mathrm{V-D}}=-0.181$~eV).
Therefore, the binding of a vacancy to a dislocation seems to be quite independent
on the dislocation character.

Looking at the variation of the interaction energy in the plane just above the glide
plane in the compressive region (Fig.~\ref{fig:vac_dislo_inter_linescan}),
one can see that the binding in the stacking fault is smaller than the one 
in the core of the partial dislocations.
We do not observe a repulsion in the middle of the stacking fault as observed 
by Häkkinen \etal \cite{HAK90}.
Thus a Suzuki segregation of vacancies is expected. 
Nevertheless, this large difference
of the binding energies indicates that vacancies cannot easily migrate from one partial 
dislocation to the other as they have to cross the stacking fault.
In such a case, one expects that diffusion is faster along the partial dislocation than in-between.
This was shown by Fang and Wang \cite{FAN00} who used the same potential
to calculate migration energies of a vacancy in Al.
They obtained a higher energy for transverse core diffusion
than for longitudinal core diffusion.
Therefore, fast diffusion should occur along two well-separated pipes corresponding
to the partial dislocations.
Pipe diffusion in a larger zone corresponding to the two pipes linked by a ribbon,
as observed by Huang \etal \cite{HUA89,HUA91} in their molecular dynamic simulations,
should only occur at high temperatures, where the partial dislocation cores 
spread over the stacking fault.

\begin{figure}[!bp]
	\begin{center}
		\includegraphics[width=0.49\linewidth]{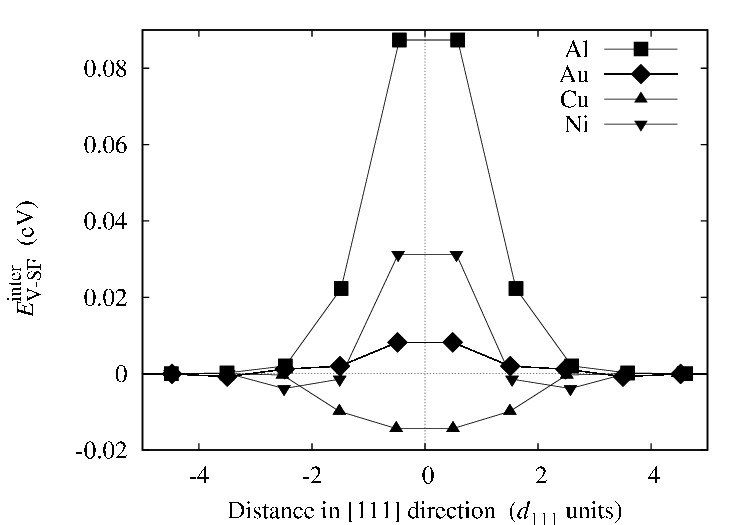}
		\hfill
		\includegraphics[width=0.49\linewidth]{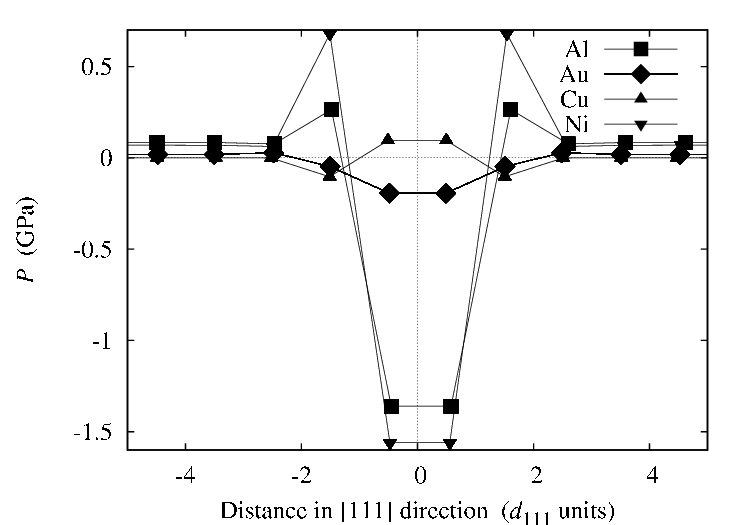}
	\end{center}
	\caption{Vacancy-stacking fault interaction energy $E^{\mathrm{inter}}_{\mathrm{V-SF}}$ (left)
	and atomic pressure $P$ (right)
	as functions of distance from the stacking fault. Distances are normalized by $d_{111}$,
	the interplanar distance in [111] direction in each material.}
	\label{fig:vac_stacking_inter}
\end{figure}

Below the glide plane, in the tensile zone, most of the vacancy 
positions are repulsive and only a few ones are slightly attractive. 
This contrasts with first atomic calculations
which were performed with pair potentials and predicted an attraction 
of the vacancy above and below the glide plane \cite{PER72,YAK85}.

Concerning the configuration in which the repulsive interaction is maximum,
one cannot draw a general situation.
Indeed, for Au and Cu this position is located above the core of the partial dislocations,
whereas for Al and Ni it is located in-between.
Moreover, for Al and Cu, it lies three planes below the glide plane, 
for Au two planes below and for Ni only one plane below,
\ie in the habit plane of the stacking fault.
We try to rationalize these differences by looking at the interaction energy 
$E^{\mathrm{inter}}_{\mathrm{V-SF}}$ between the vacancy and 
the stacking fault (Fig.~\ref{fig:vac_stacking_inter}).
When the vacancy lies in the habit plane of the stacking fault,
this interaction is repulsive for Al and Ni,
repulsive too for Au, but an order of magnitude lower,
and attractive for Cu.
This opposite interaction in the case of Cu arises from the fact that 
the stacking fault in Cu is in compression, whereas it is in tension in Al, Au and Ni.
This was seen by computing the atomic pressure from the EAM potential using 
the definition of Vitek and Egami \cite{VIT87} (Fig.~\ref{fig:vac_stacking_inter}).
This repulsive interaction with the stacking fault in Al and Ni adds,
in the tensile zone, to the elastic repulsive interaction with the partial dislocations. 
Therefore the strongest repulsion is found between the partial dislocations.
On the other hand, in the case of Cu, there is a compensation between the interactions
with the stacking fault and the partial dislocations.
Nevertheless, this explanation which tries to split the interaction in two different contributions
does not really allow to predict the binding or the repulsion of the vacancy close to the 
dislocation. This shows the usefulness of an atomic calculation close
to the dislocation cores where the vacancy behavior is hard to predict.

\section{Elasticity}

As long as the vacancy is far enough from the dislocation,
the interaction energy between both defects can be predicted 
using the elasticity theory. Considering only the first-order size interaction, 
this energy is simply given by \cite{BUL70,COT49,FRI64,HIR82} 
\begin{equation}
	\label{eq:Einter_elas1}
	E^{\mathrm{inter}}_{\mathrm{V-D}} = P\ \delta\Omega_{\mathrm{V}},
\end{equation}
where $P=-\left( \sum_{ii}\sigma_{ii} \right)/3$ is the pressure 
created by the dislocations at the lattice site where the vacancy lies 
and $\delta\Omega_{\mathrm{V}}$ is the vacancy relaxation volume. 
This last term is calculated directly from the atomic potential
whereas the pressure is computed within the framework of the elasticity theory
of dislocations \cite{HIR82}.
Usually such calculations are performed assuming perfect dislocations
and isotropic elasticity. However, dislocations are dissociated in fcc metals
and not all materials can be assumed isotropic.
We therefore check the validity of these different simplifying assumptions.
For the isotropic elasticity calculations, we use the shear modulus $\mu$, the bulk modulus $K$ 
and the Poisson coefficient $\nu$, obtained by Voigt average \cite{HIR82} of the elastic constants
corresponding to the potential used for atomic calculations:
\begin{subequations}
	\label{eq:elastic_Voigt}
	\begin{align}
	\mu &= \frac{1}{5} \left( C_{11} - C_{12} + 3C_{44} \right), \\
	K &= \frac{1}{3}\left( C_{11}+2C_{12} \right), \\
	\nu &= \frac{C_{11} + 4C_{12} - 2C_{44}}{2\left( 2C_{11} + 3C_{12} + C_{44} \right)}.
	\end{align}
\end{subequations}

Whatever the assumptions used to calculate the pressure within the elasticity theory
of dislocations, one has to cope with the problem of the periodic boundary conditions
corresponding to the atomic simulations. Indeed, the two dislocations 
in the simulation box are not the only ones to consider in the elastic calculation.
All their periodic images forming an infinite array of dipoles have to be 
included too.
As the stress created by a dislocation is varying as the inverse of the 
distance, the sum associated with the calculation of the pressure
is only conditionally convergent. To circumvent this numerical problem,
we use the computational approach proposed by Cai\footnote{So as to compute the stress
field created by this double periodic array of dislocations, one can take advantage too
from the fact that each dislocation and its periodic images in the z direction form
a tilt boundary. This causes a stress field which expression is analytical \cite{HIR82}
and is known to decay exponentially with the distance normal to the wall.} \cite{CAI05,CAI03}.

\subsection{Perfect dislocations within the isotropic elasticity theory}

\begin{figure}[!bp]
	\begin{center}
		\subfigure[{$[110]$ direction.}]{
		\includegraphics[width=0.47\linewidth]{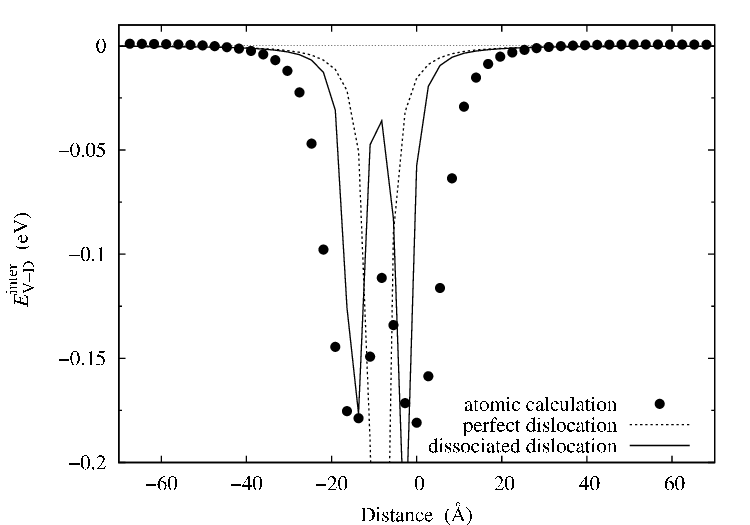}}
		\hfill
		\subfigure[{$[\bar{1}11]$ direction.}]{
		\includegraphics[width=0.47\linewidth]{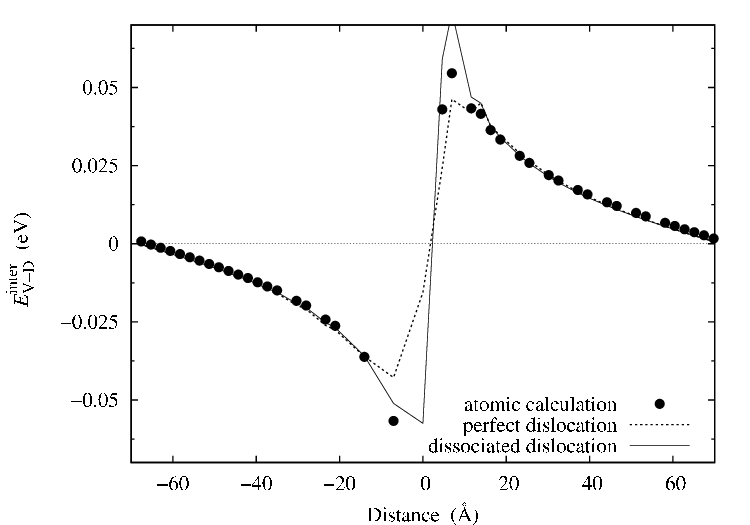}}
	\end{center}
	\caption{Variation with the separation distance of the vacancy - dislocation interaction energy 
	$E^{\mathrm{inter}}_{\mathrm{V-D}}$ in $[110]$ and $[\bar{1}11]$ directions in Al.
	Symbols correspond to atomic simulation and lines to the isotropic elasticity theory 
	(Eq.~\ref{eq:Einter_elas1}) assuming perfect or dissociated dislocations.}
	\label{fig:Einter_dissociation}
\end{figure}

Assuming isotropic elasticity, the pressure created by a perfect edge dislocation
at a point of coordinates $(x,y,z)$\footnote{The origin is defined as the 
center of the perfect dislocation and axes are defined on Fig.~\ref{fig:box_sketch}a.}
is
\begin{equation}
	\label{eq:P_isotropic_perfect}
	P_i = \delta_i \frac{\mu b}{3\pi} \frac{1+\nu}{1-\nu} \frac{z}{x^2+z^2},
\end{equation}
where $\delta_{i}=+1$ (conversely $-1$) if the two supplementary half-plane corresponding
to the edge dislocation $i$ is in the $z>0$ (conversely $z<0$) sub-space and $b=a\sqrt{2}/2$
is the Burgers vector of the perfect edge dislocation.
This pressure, predicted by isotropic elasticity, can be used in Eq.~\ref{eq:Einter_elas1}
after summing pressures arising from all image dislocations so as to predict the interaction energy 
between the vacancy and the dislocations.
In Fig.~\ref{fig:Einter_dissociation}, we compare the elasticity theory predictions 
with values of $E^{\mathrm{inter}}_{\mathrm{V-D}}$
previously calculated with EAM potentials, for different separation distances between both defects
in the $[110]$ and $[\bar{1}11]$ directions.
The elasticity theory manages to follow the atomic calculations 
as long as the vacancy is not too close to the dislocation core.
For Al, a reasonable agreement is obtained for distances
greater than $\sim10$~\AA{ }in $[\bar{1}11]$ direction 
and greater than $\sim20$~\AA{ }in $[110]$ direction.
We believe that the elasticity theory is less accurate
when the pressure is close to zero.

\subsection{Influence of dislocations dissociation}

We then examine whether considering the dissociation 
of the dislocations modifies the value of the interaction energy with the vacancy.
According to isotropic elasticity theory, the dissociation distance is\cite{FRI64,HIR82}
\begin{equation}
	\label{eq:dissociation_iso}
	d^{\mathrm{iso}} = \frac{\mu b^2}{24\pi\gamma}\frac{2+\nu}{1-\nu},
\end{equation}
As presented in table \ref{tab:para}, the dissociation distance predicted 
is quite close to the one observed in atomic simulations.

Only the edge components of each partial dislocations contribute to the pressure
which bewomes
\begin{equation}
	\label{eq:P_isotropic_dissociated}
	P_i = \delta_i \frac{\mu b}{6\pi} \frac{1+\nu}{1-\nu} \left( 
	\frac{z}{(x-d^{\mathrm{iso}})^2+z^2} + \frac{z}{(x+d^{\mathrm{iso}})^2+z^2} \right).
\end{equation}
When the vacancy is close to the dislocation, the interaction energy 
$E^{\mathrm{inter}}_{\mathrm{V-D}}$ calculated using the pressure,
as given by Eq.~\ref{eq:P_isotropic_dissociated},
is in better agreement with atomic calculations than with the pressure 
given by Eq.~\ref{eq:P_isotropic_perfect}.
Even for Al, which has the smallest dissociation distance, the improvement is obvious 
(Fig.~\ref{fig:Einter_dissociation}).
When the separation distance between the vacancy and the dislocation becomes large
compared to the dissociation distance ($\sqrt{x^2+z^2} >> d^{\mathrm{iso}}$), 
the difference between the dissociated and the non-dissociated cases becomes negligible
since Eq.~\ref{eq:P_isotropic_dissociated} converges to Eq.~\ref{eq:P_isotropic_perfect}.

\subsection{Influence of anisotropy}

\begin{figure}[!bp]
	\begin{center}
		\subfigure[{$[110]$ direction.}]{
		\includegraphics[width=0.47\linewidth]{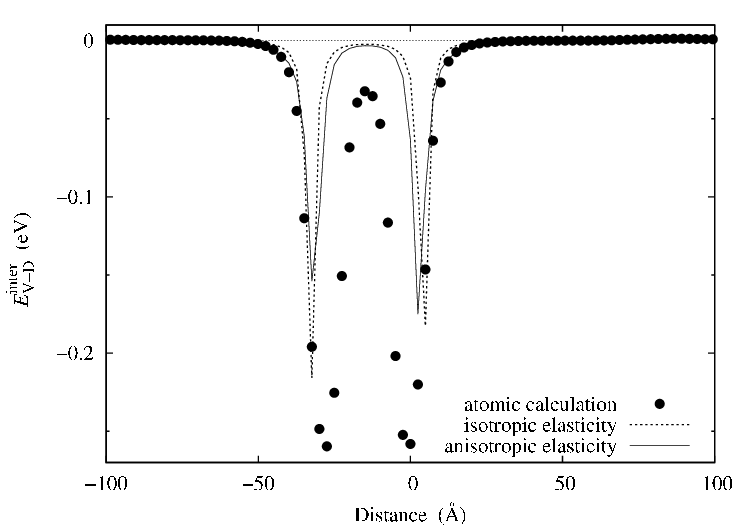}}
		\hfill
		\subfigure[{$[\bar{1}11]$ direction.}]{
		\includegraphics[width=0.47\linewidth]{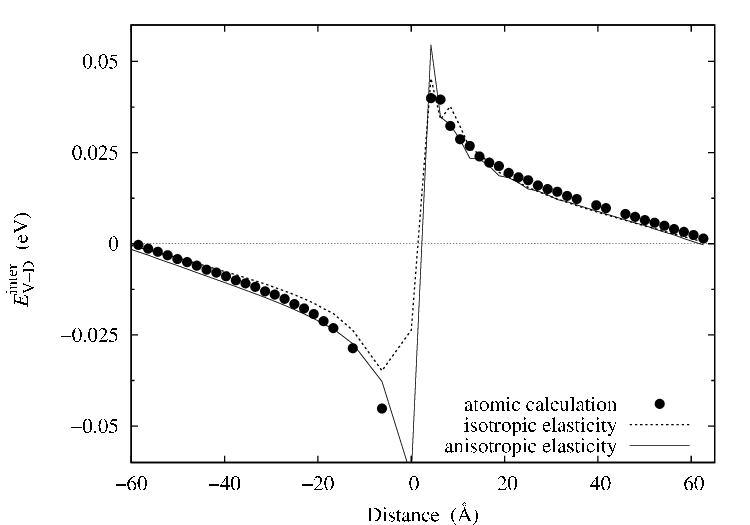}}
	\end{center}
	\caption{Variation with the separation distance of the vacancy - dislocation interaction energy 
	$E^{\mathrm{inter}}_{\mathrm{V-D}}$ in $[110]$ and $[\bar{1}11]$ directions in Cu.
	Symbols correspond to atomic simulation and lines to the isotropic or anisotropic elasticity theory 
	(Eq.~\ref{eq:Einter_elas1}) assuming dissociated dislocations.}
	\label{fig:Einter_isotropy}
\end{figure}

Instead of assuming isotropic elasticity to calculate the pressure created by a dislocation,
one can use the elastic constants $C_{11}$, $C_{12}$ and $C_{44}$ corresponding to the empirical
potential of the atomic simulations and take full account of the anisotropy. 
Considering anisotropy can be quite significant in elastic calculations. 
For instance, two dilatation centers do not interact within isotropic elasticity,
whereas there is an interaction when anisotropy is introduced.
Moreover, Cai \cite{CAI05,CAI03} showed that anisotropy has to be considered when calculating 
dislocation basic properties like their core energy or the Peierls stress.

Due to the symmetry in the $[1\bar{1}2]$ direction along the dislocation line, 
finding the stress field associated with an edge dislocation is a two-dimensional 
problem and can be solved numerically following the sextic anisotropic elasticity 
theory of straight dislocations \cite{HIR82}.
We thus calculate the elastic interaction between the two partial dislocations composing
an edge dislocation so as to obtain its dissociation distance $d^{ani}$.
The obtained value is really close to the one calculated with isotropic elasticity (Tab.~\ref{tab:para}).
The agreement with the distance observed in atomic simulation is slightly better but 
the difference is not really relevant, even for Cu which is the most anisotropic metal considered.

Using anisotropic elasticity theory to compute the pressure 
created by the dissociated edge dislocations,
Eq.~\ref{eq:Einter_elas1} gives another estimate for the interaction energy with vacancies. 
As shown for Cu in Fig.~\ref{fig:Einter_isotropy}, the values of $E^{\mathrm{inter}}_{\mathrm{V-D}}$
are similar to the ones calculated with isotropic elasticity. 
The agreement with atomic calculations is slightly better, but, as for the dissociation distance, 
the change due to anisotropy is negligible. 
Anisotropy does not lead to any significant change compared to isotropic calculations. 
Therefore, it does not need to be considered in the elastic calculation of the interaction energy
of a vacancy with dislocations.

\subsection{Inhomogeneity interaction}

\begin{figure}[!bp]
	\begin{center}
		\subfigure[{$[110]$ direction.}]{
		\includegraphics[width=0.47\linewidth]{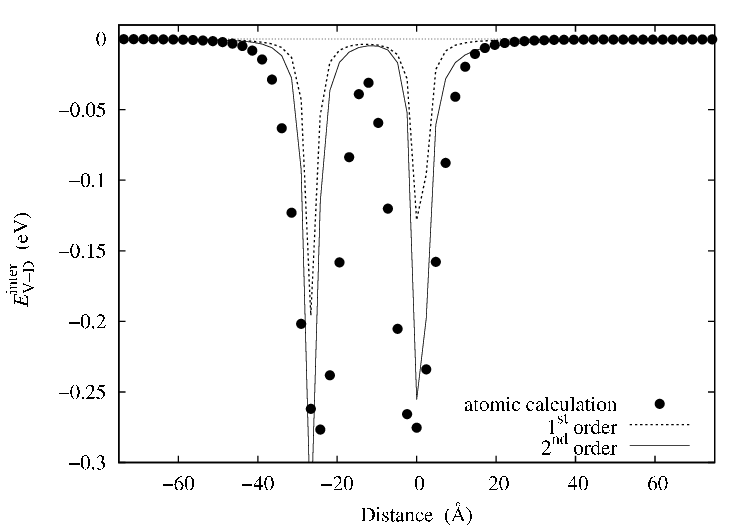}}
		\hfill
		\subfigure[{$[\bar{1}11]$ direction.}]{
		\includegraphics[width=0.47\linewidth]{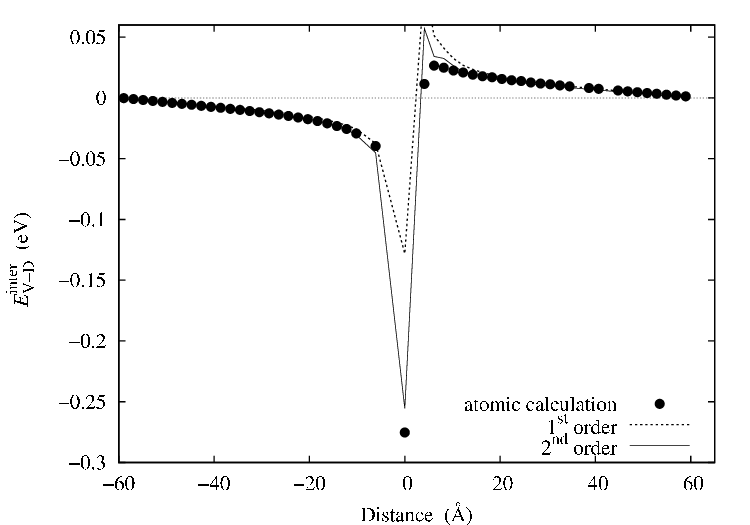}}
	\end{center}
	\caption{Variation with the separation distance of the vacancy - dislocation interaction energy 
	$E^{\mathrm{inter}}_{\mathrm{V-D}}$ in $[110]$ and $[\bar{1}11]$ directions in Ni.
	Symbols correspond to atomic simulation and lines to the isotropic elasticity theory 
	assuming dissociated dislocations.
	The first order approximation corresponds to Eq.~\ref{eq:Einter_elas1}
	and the second order to Eq.~\ref{eq:Einter_elas2}.}
	\label{fig:Einter_ordre}
\end{figure}

Another elastic interaction between the vacancy and the dislocation, 
the inhomogeneity contribution, needs to be considered \cite{BUL70,BUL63}. 
It arises from the fact that the vacancy has different elastic constants from the matrix.
Thus, there is a change in the elastic energy that
can be stored in the material.
As the elastic constants that can be associated to the vacancy have to be lower
than the ones of the matrix, this inhomogeneity interaction is always attractive.
Indeed, the presence of a vacancy releases some elastic energy. 
Assuming that the elastic constants of the vacancy are null, one maximizes 
this contribution. 
In isotropic elasticity, the interaction energy between the vacancy and the dislocations
now reads 
\begin{equation}
	\label{eq:Einter_elas2}
	E^{\mathrm{inter}}_{\mathrm{V-D}} = P\ \delta\Omega_{\mathrm{V}}
	-\frac{1}{2}\Omega \left( \frac{3(1-\nu)}{2(1-2\nu)K}P^2
	+ \frac{5(1-\nu)}{(7-5\nu)\mu} \bar{\sigma}^2 \right),
\end{equation}
where $\bar{\sigma}$ is the Von-Misès equivalent stress and is related 
to the pure shear components of the stress created by the dislocation by
\begin{equation}
	\label{eq:sigmaVM}
	\bar{\sigma} = \sqrt{\frac{3}{2}\sum_{i,j}\left( \sigma_{ij}+P\delta_{ij} \right)^2}.
\end{equation}
Like the pressure, it can be easily computed 
for dissociated edge dislocations using isotropic elasticity.
Considering the inhomogeneity contribution, the interaction energy now depends 
not only on the pressure but also on the shear stress.
As it is varying linearly with the square of the stress, and thus with the square
of the inverse of the distance between both defects, this contribution is much 
shorter range than the size contribution.

Considering the inhomogeneity contribution in the isotropic elasticity calculation of $E^{inter}_{\mathrm{V-D}}$
improves the agreement with atomic simulations close to the core of the partial dislocations,
as can be seen for Ni in Fig.~\ref{fig:Einter_ordre}. In particular, without this contribution
the interaction is symmetric with respect to the dislocation glide plane.
The inhomogeneity interaction thus explains
why the vacancy attraction in the compressive region is larger in magnitude
than its repulsion in the tensile zone.
This non-symmetric behavior of the interaction energy is clearly visible 
when the vacancy lies no more than one atomic distance from the dislocation core
(Fig.~\ref{fig:Einter_ordre}(b)).
For larger separation distances, the change induced by the inhomogeneity contribution 
is not significant.
As pointed by Bullough and Newman \cite{BUL70}, when this inhomogeneity interaction
is taken into account in the elastic calculation, one should incorporate also
the second-order size interaction arising from the non-linear elastic properties 
of the crystal, as both contributions are decreasing in the same manner with the distance.
Eventhough this last term may lead to a better 
agreement with atomic calculations in the vicinity of the dislocation cores,
we do not consider it since results based on non-linear elastic calculations
are too complicated to handle.
Therefore, we prefer to use only the linear elasticity theory, considering the fact
that one can switch to atomic calculations for short separation distances 
where the theory breaks down.

\section{Conclusions}

The interaction between vacancies and edge dislocations has been 
modeled in fcc metals at different scales.
When only a few atomic distances lie between the vacancy and the cores of the partial dislocations
or the stacking fault an atomic description of this interaction is needed.
Using empirical potentials, one finds for fcc metals that the most attractive position 
of the vacancy lies at the edges of the supplementary half-planes
corresponding to the partial dislocations.
Positions above the slip planes, in the compressive region, are always attractive
whereas, below the slip planes, positions are repulsive or, for some metals like Cu and Ni,
slightly attractive in the habit plane of the stacking fault.

When the separation distance between both defects is greater, the elasticity theory 
manages to predict quantitatively the interaction energy. 
One does not need to take into account anisotropy in the calculations
as this does not lead to any significant change compared to an isotropic calculation.
On the other hand, better agreement with atomic simulations is obtained
for short separation distances, when the dislocation dissociation is considered.
Taking into account the inhomogeneity interaction and not only the size effects
improves predictions of the elastic calculation too.
The most significant change is that the interaction energy is not anymore symmetric with respect
to the dislocation glide plane.

\begin{ack}
	The author is grateful to Dr. A. Barbu, Dr. J.-L. Bocquet, Dr. M. Guttmann, 
	Dr. B. Legrand, Dr. D. Mordehai, and Prof. J.-P. Poirier for fruitful discussions.
\end{ack}

\bibliographystyle{elsart-num}
\bibliography{clouet2006}

\end{document}